\documentclass[twocolumn,showpacs,preprintnumbers,amsmath,amssymb]{revtex4}
\usepackage{graphicx}% Include figure files
\usepackage{dcolumn}% Align table columns on decimal point
\usepackage{bm}% bold math

\usepackage{amssymb}
\usepackage{amsmath}

\begin{document}

%%\preprint{for Phys. Rev. D (\textbf{Ver.06}) }

\title{General form of entropy on the horizon of the universe in entropic cosmology}

\author{Nobuyoshi {\sc Komatsu}$^{1}$}  \altaffiliation{E-mail: komatsu@se.kanazawa-u.ac.jp} 
\author{Shigeo     {\sc Kimura}$^{2}$}

\affiliation{$^{1}$Department of Mechanical Systems Engineering, Kanazawa University, 
                          Kakuma-machi, Kanazawa, Ishikawa 920-1192, Japan \\
                $^{2}$The Institute of Nature and Environmental Technology, Kanazawa University, 
                          Kakuma-machi, Kanazawa, Ishikawa 920-1192, Japan}%
\date{\today}

\begin{abstract}
Entropic cosmology assumes several forms of entropy on the horizon of the universe, where the entropy can be considered to behave as if it were related to the exchange (the transfer) of energy. To discuss this exchangeability, the consistency of the two continuity equations obtained from two different methods is examined, focusing on a homogeneous, isotropic, spatially flat, and matter-dominated universe. The first continuity equation is derived from the first law of thermodynamics, whereas the second equation is from the Friedmann and acceleration equations. To study the influence of forms of entropy on the consistency, a phenomenological entropic-force model is examined, using a general form of entropy proportional to the $n$-th power of the Hubble horizon. In this formulation, the Bekenstein entropy (an area entropy), the Tsallis--Cirto black-hole entropy (a volume entropy), and a quartic entropy are represented by $n=2$, $3$, and $4$, respectively. The two continuity equations for the present model are found to be consistent with each other, especially when $n=2$, i.e., the Bekenstein entropy. The exchange of energy between the bulk (the universe) and the boundary (the horizon of the universe) should be a viable scenario consistent with the holographic principle.
\end{abstract}

%%\keywords{Cosmology}
\maketitle

\section{Introduction}

The accelerated expansion of the late universe \cite{PERL1998ab_Riess1998_2004} can be elegantly explained by $\Lambda$CDM (lambda cold dark matter) models that assume a cosmological constant $\Lambda$ and dark energy. However, it is well-known that the $\Lambda$CDM model suffers from theoretical difficulties, such as the cosmological constant problem \cite{Weinberg1989}. To resolve the difficulties, $\Lambda (t)$CDM models, in which a time-varying cosmological term $\Lambda (t)$ is assumed \cite{Freese-Mimoso,Waga1994_Sola_2007-2009,Sola_2009-2015,Sola_2013b,LimaSola_2013a,Valent2015,Sola_2015_1c,Sola_2015_1a,Sola_2015_1b}, have been extensively examined. (For various other models, see, e.g., Refs.\ \cite{Miao1,Bamba1,Weinberg1,Roy1} and references therein.)

As a possible alternative scenario, entropic-force models based on the holographic principle \cite{Hooft-Bousso}, in which several forms of entropy on the horizon of the universe are assumed \cite{Easson12,Cai1,Koivisto-Costa1,Lepe1,Basilakos1,Sola_2014a,Koma4,Koma5,Koma6,Koma7,Koma8,Gohar_2015a,Gohar_2015b,Nunes_2015b}, have been recently proposed. For example, the Bekenstein entropy (an area entropy based on additive statistics) \cite{Bekenstein1}, the Tsallis--Cirto entropy (a volume entropy based on nonadditive statistics) \cite{Tsallis2012}, and a quartic entropy \cite{Koma5,Koma6} have been suggested for the entropy on the horizon \cite{Easson12,Koma6}. Most entropic-force models can be interpreted as a particular case of $\Lambda (t)$CDM models \cite{Basilakos1,Sola_2014a}. This interpretation implies that the assumed entropy is exchangeable (reversible), such as is the entropy related to the exchange of energy \cite{Prigogine_1998}.  That is, the entropy on the horizon is considered to behave as if it were related to  `energy exchange cosmology', which assumes the transfer of energy between two fluids \cite{Barrow22}, e.g., the interaction between dark matter and dark energy, dynamical vacuum energy, etc. \cite{Wang0102_YWang2014,Pavon_2005}.

Such pairs of fluids are not generally employed in entropic cosmology because dark energy is not assumed. Accordingly, the exchangeability may imply the transfer of energy between the bulk (the universe) and the boundary (the horizon of the universe) \cite{Lepe1}, because the information of the bulk is assumed to be holographically stored in the boundary \cite{Easson12}. However, the exchangeability has not yet been made clear in entropic cosmology. The exchangeability can probably be discussed in terms of the consistency of two continuity equations derived from two different methods \cite{Koma4}. For example, the continuity equation is typically derived from the Friedmann and acceleration equations because only two of the three equations are independent \cite{Ryden1}. Alternatively, the continuity equation can be derived from the first law of thermodynamics as well \cite{Koma4,Ryden1}. Therefore, it is possible to discuss the consistency of the continuity equations derived from these two different methods. The forms of entropy on the horizon (i.e., area, volume, and quartic entropies) are expected to affect the consistency.

In contrast, several entropic-force models similar to bulk viscous models \cite{Weinberg0,Murphy1,Barrow1986-Lima1988,Zimdahl1996-Nojiri2011} and CCDM (creation of cold dark matter) models \cite{Lima_1992b,Lima_1992e,Lima_2000,Lima_2014b,Harko_2014,Zimdahl_1993,Lima-Others1996-2008,Lima2010-2014,Lima_Newtonian_1997,Lima2011,Ramos_2014-2014b,C301} assume irreversible entropy related to dissipation processes \cite{Koma5,Koma6}.
In those models, an effective description for pressure is used without assuming the exchange of energy. However, Prigogine \textit{et al.} have proposed open systems with the exchange of energy, in which reversible and irreversible entropies are considered \cite{Prigogine_1988b,Prigogine1989}, to discuss the thermodynamics of cosmological matter creation for non-adiabatic processes. The proposed system is suitable for describing the general systems in entropic cosmology discussed here.

In this context, we formulate a phenomenological entropic-force model, in which area, volume, and quartic entropies \cite{Koma4,Koma5,Koma6} are systematically assumed to be the entropy on the horizon. Moreover, irreversible entropy due to matter creation \cite{Koma7,Koma8} is included in that formulation. Using the present model, we examine whether the entropy on the horizon behaves exchangeably or not. In this short paper, to discuss the exchangeability, we focus on the consistency of the two continuity equations derived from different methods. The study of the exchangeability should provide new insights into entropic cosmology.

The remainder of the article is organized as follows. In Sec.\ \ref{General entropic-force models}, a phenomenological entropic-force model is formulated, assuming a general form of entropy on the horizon. In Sec.\ \ref{Two continuity equations}, two continuity equations are derived from two different methods. Specifically, in Sec.\ \ref{Continuity equation from the first law of thermodynamics}, the continuity equation is derived from the first law of thermodynamics, and in Sec.\ \ref{Continuity equation from the Friedmann equations}, the continuity equation is derived from the Friedmann and acceleration equations. The consistency of the two continuity equations is then discussed in Sec.\ \ref{Consistency}. Finally, in Sec.\ \ref{Conclusions}, the conclusions of the study are presented.

\section{Entropic-force models} 
\label{General entropic-force models}

In this section, a phenomenological entropic-force model that assumes a general form of entropy on the horizon of the universe is described. For this purpose, a homogeneous, isotropic, and spatially flat universe is considered, and the scale factor $a(t)$ is examined at time $t$ in the Friedmann--Lema\^{i}tre--Robertson--Walker metric \cite{Koma4,Koma5,Koma6,Koma7,Koma8}. First, $\Lambda$CDM models are briefly reviewed in Sec.\ \ref{LCDM models}. The entropic-force model is then formulated in Sec.\ \ref{Entropic-force model based on a general form of entropy}. The derivation of entropic forces  is based on the original work of Easson \textit{et al.} \cite{Easson12} and the recent work of the present authors \cite{Koma4,Koma5,Koma6}. The concept of entropic force considered here is different from the idea  that gravity itself is an entropic force \cite{Padma1,Verlinde1}, as described in Ref.\ \cite{Easson12}. In the present paper, the inflation of the early universe is not discussed because we have chosen to focus on background evolution of the late universe.

Please note that irreversible entropy due to matter creation is \textit{not} considered in this section. The irreversible entropy is discussed in the next sections.

\subsection{$\Lambda$CDM model} 
\label{LCDM models}

In this subsection, the well-known $\Lambda$CDM models are briefly reviewed \cite{Ryden1,Weinberg1}. The acceleration equation is written as
\begin{equation}
  \frac{ \ddot{a}(t) }{ a(t) }   =  \dot{H}(t) + H(t)^{2}   
                                          =  -  \frac{ 4\pi G }{ 3 }  \left (  \rho(t) +  \frac{3p(t)}{c^2}  \right )  + \frac{\Lambda}{3}  ,
\label{eq:FRW2_LCDM}
\end{equation}
where the Hubble parameter $H(t)$ is defined by
\begin{equation}
   H(t) \equiv   \frac{ da/dt }{a(t)} =   \frac{ \dot{a}(t) } {a(t)}  ,
\label{eq:Hubble}
\end{equation}
and $G$, $\Lambda$, $c$, $\rho(t)$, and $p(t)$ are the gravitational constant, a cosmological constant, the speed of light, the mass density of cosmological fluids, and the pressure of cosmological fluids, respectively \cite{Koma5}. The right-hand side of Eq.\ (\ref{eq:FRW2_LCDM}) includes a driving term $\Lambda/3$, which can explain the accelerated expansion of the late universe. This term corresponds to a cosmological constant term and is interpreted as an additional energy component called dark energy.

\subsection{Entropic-force model based on a general form of entropy proportional to $r_{H}^{n}$} 
\label{Entropic-force model based on a general form of entropy} 

In entropic-force models, extra driving terms are derived from entropic forces, unlike in $\Lambda$CDM models \cite{Easson12}. The entropic-force model assumes that the horizon of the universe has an associated entropy $S$ and an approximate temperature $T$. In this study, we use the Hubble horizon as the preferred screen because the apparent horizon coincides with the Hubble horizon in a spatially flat universe \cite{Easson12}. If we were instead considering a spatially non-flat universe, we would use the apparent horizon as the preferred screen \cite{Easson12}.

The Hubble horizon (radius) $r_{H}$ is given by
\begin{equation}
     r_{H} = \frac{c}{H}  \quad  \textrm{and therefore}    \quad    \dot{r}_{H}  =  - \frac{ H  \dot{H} }{ c^2 } r_{H}^3     .
\label{eq:rH}
\end{equation}
The temperature $T$ on the horizon is given by 
\begin{equation}
 T = \frac{ \hbar H}{   2 \pi  k_{B}  } \times   \gamma  =  \frac{ \hbar }{   2 \pi  k_{B}  }  \frac{c}{ r_{H} }   \gamma    ,   
\label{eq:T0}
\end{equation}
where $k_{B}$ and $\hbar$ are the Boltzmann constant and the reduced Planck constant, respectively. 
It should be noted that the temperature considered here is assumed to be obtained by multiplying the horizon temperature, $ \hbar H /( 2 \pi k_{B} ) $, by a non-negative free parameter, $\gamma$.
Here, $\gamma$ is assumed to be a free parameter for the temperature \cite{Koma4,Koma5}.

In the present study, we do not discuss the magnitude of the free parameter $\gamma$ for the temperature. 
However, before proceeding further, in this paragraph, we give a brief review of $\gamma$, according to the previous and recent studies.
For example, Easson \textit{et al.} have suggested a similar modified coefficient for the temperature, in which $\gamma$ may be estimated from a derivation of surface terms or the Hawking temperature description \cite{Easson12}. 
Also, Cai \textit{et al.} have proposed that $\gamma$ can be interpreted as a parameter for the holographic screen temperature \cite{Cai1}. 
In those works, $\gamma$ is considered to be of the order of $O(1)$ \cite{Koma4}.
On the other hand, interestingly, D\c{a}browski \textit{et al.} have recently reported that a similar parameter $\gamma$ is two to four orders of magnitude less than $O(1)$ \cite{Gohar_2015b}.
In that paper, the combination of the holographic and vacuum dark energies is likely assumed from different viewpoints.
Therefore, $\gamma$ used in Ref.\ \cite{Gohar_2015b} should be related to a parameter $\nu$ on dynamical vacuum models (see the second paper of Ref.\ \cite{Valent2015}). 
The parameter $\nu$ can be small because it behaves as a type of $\beta$-function coefficient in quantum field theory.
In fact, $\nu$ is expected to be approximately $\nu \sim 10^{-3}$ from observations, as examined by Sol\`{a} \textit{et al.} \cite{Sola_2015_1c}.
Consequently, a similar parameter $\gamma$ used in Ref.\ \cite{Gohar_2015b} may become small as well.
The similar parameter is expected to be related to $\gamma$ considered here. 
The discussion of $\gamma$ will provide new insights in entropic cosmology because the smallness of $\gamma$ has not yet been explained by the holographic approach. 
This task is left for the future research. 
Keep in mind that $\gamma$ considered here is assumed to be a free parameter.

In the original entropic-force model \cite{Easson12}, an associated entropy on the Hubble horizon is given as
\begin{equation} 
S_{r2}  = \frac{ k_{B} c^3 }{  \hbar G }  \frac{A_{H}}{4} =   \frac{ k_{B} c^3 }{  \hbar G }  \frac{ 4 \pi r_{H}^2 }{4} =  \frac{ \pi k_{B} c^3 }{  \hbar G }  r_{H}^2       ,
\label{eq:SH(r2)}
\end{equation}
where $A_{H}$ is the surface area of a sphere with the Hubble radius $r_{H}$. This entropy is the Bekenstein entropy proportional to area and $r_{H}^{2}$ \cite{Bekenstein1}. Recently, several forms of entropy have been proposed for the entropy on the horizon of the universe. For example, a volume entropy $S_{r3}$ and a quartic entropy $S_{r4}$ (proportional to $r_{H}^{4}$) have been used for entropic-force models \cite{Koma5,Koma6}. The volume entropy $S_{r3}$ is a generalized black-hole entropy, i.e., the Tsallis--Cirto black-hole entropy \cite{Tsallis2012}, based on nonadditive statistics \cite{Tsa0}. In contrast, although the meaning of $S_{r4}$ is less clear, it can be considered as a form of entropy that would arise if extra dimensions existed \cite{Koma6}. Consequently, it is found that an area entropy $S_{r2}$, a volume entropy $S_{r3}$, and a quartic entropy $S_{r4}$ can lead to $H^{2}$, $H$, and constant entropic-force terms, respectively. Each entropic-force term has been separately discussed in Ref.\ \cite{Koma6}. In the present study, a general form of entropy is used to discuss a phenomenological entropic-force model systematically. Note that D\c{a}browski, Gohar, and Salzano have recently proposed more extended entropic forces to examine varying-constant theories \cite{Gohar_2015a,Gohar_2015b}.

The general form of entropy (proportional to $r_{H}^{n}$) is defined by 
\begin{equation}
S_{rn} =  \frac{  \pi  k_{B} c^3 }{  \hbar G } \times L_{n}  r_{H}^{n}      \quad (n=2, 3, 4)     ,
\label{eq:S(rn)}
\end{equation}
where $n=2$, $3$, and $4$ correspond to indices of area, volume, and quartic entropies, respectively. $L_{n}$ is a non-negative free-constant-parameter. The following derivation can be applied to higher-order forms of entropy. (Values of $ L_{2}=1$, $ L_{3}=\zeta$, and $ L_{4}=\psi$ were used for the free-constant parameters in Ref.\ \cite{Koma6}.)

We now derive an entropic-force $F_{rn}$ from a general form of entropy, $S_{rn} \propto r_{H}^{n}$. The entropic force can be given by 
\begin{equation}
    F_{rn}  =  -  \frac{dE}{dr}  =  - T \frac{dS_{rn}  }{dr}   \left ( =    - T \frac{dS_{rn}  }{dr_{H}} \right )        ,     
\label{eq:Frn}
\end{equation}
where the minus sign indicates the direction of increasing entropy or the screen corresponding to the horizon \cite{Easson12}.  Substituting  Eqs.\ (\ref{eq:T0}) and (\ref{eq:S(rn)}) into Eq.\ (\ref{eq:Frn}), the entropic-force $F_{rn}$ becomes 
\begin{align}
    F_{rn}  &=      - T \frac{dS_{rn}}{dr_{H}}           
                =  -  \frac{ \hbar }{ 2 \pi k_{B}  }  \frac{c}{ r_{H} }  \gamma   \times   \frac{d}{dr_{H}}  \left [ \frac{  \pi  k_{B} c^3 }{  \hbar G }  \times L_{n}  r_{H}^{n}    \right ]     \notag \\
           &=  -  \gamma  \frac{c^{4}}{G} \left ( \frac{n L_{n} }{2} \right )  r_{H}^{n-2}      .
\label{eq:F(rn)}
\end{align}
From Eq.\ (\ref{eq:F(rn)}), the pressure $p_{rn}$ is given as 
\begin{align}
   p_{rn}  &=    \frac{ F_{rn} } {A_{H}}   =  -  \gamma  \frac{c^{4}}{G} \left ( \frac{n L_{n} }{2} \right )  r_{H}^{n-2}     \frac{1} {4 \pi r_{H}^2}     \notag \\ 
                     &=  -  \gamma  \frac{c^{4}  n L_{n} }{  8 \pi G }   r_{H}^{n-4}                                                                  
                       =  -  \gamma  \frac{c^{4}  n L_{n} }{  8 \pi G }   \left ( \frac{c }{H} \right )^{n-4}                                                               \notag \\ 
                     &=  -  \gamma  \left ( \frac{c^{n}  n L_{n} }{  8 \pi G }  \right  )  H^{4-n}                            .      
\label{eq:P(rn)}
\end{align}
In entropic cosmology \cite{Easson12},  $p_{rn}$ is added to the acceleration equation. To this end, Eq.\ (\ref{eq:FRW2_LCDM}) is arranged as follows. Setting $\Lambda =0$, replacing $p$ by $p + p_{rn}$, and substituting Eq.\ (\ref{eq:P(rn)}) into Eq.\ (\ref{eq:FRW2_LCDM}), the acceleration equation is given by 
\begin{align}
  \frac{ \ddot{a} }{ a }  &=   -  \frac{ 4\pi G }{ 3 }  \left (  \rho +  \frac{ 3 (p+p_{rn}) }{c^2}  \right )                                                                                               \notag \\ 
                                 &=   -  \frac{ 4\pi G }{ 3 }  \left (  \rho +  \frac{ 3  p }{c^2}  \right )    +   \gamma  \left ( \frac{c^{n-2}  n L_{n} }{  2 }  \right  )  H^{4-n}      .
\label{eq:FRW2_(rn)}
\end{align}
The last term on the right-hand side is the so-called entropic-force term. As for most entropic-force models, adding the entropic-force term to the Friedmann equation $H^{2}  =    8\pi G \rho/ 3$ gives  
\begin{equation}
  H^{2}  =    \frac{ 8\pi G }{ 3 }    \rho      +   \gamma  \left ( \frac{c^{n-2}  n L_{n} }{  2 }  \right  )  H^{4-n}      .
\label{eq:FRW1_(rn)}
\end{equation}
For $n=2$, $3$, and $4$, the last terms on the right-hand side of Eq.\ (\ref{eq:FRW2_(rn)}) are $\gamma L_{2} H^2$, $\gamma (3c L_{3} /2)H$, and $\gamma (2c^{2} L_{4}) $, respectively. That is, the $H^{2}$, $H$, and constant terms are phenomenologically derived from the area, volume, and quartic entropies, respectively. This result agrees with that of Ref.\ \cite{Koma6}, in which $L_{2}=1$, $L_{3}=\zeta$, and $L_{4} = \psi$ were used.  Keep in mind that irreversible entropy due to matter creation is neglected in this section. Accordingly, the formulations of the Friedmann and acceleration equations are essentially equivalent to those of $\Lambda (t)$CDM models. This type of $\Lambda(t)$CDM model has been examined extensively (see, e.g.,  Refs.\ \cite{Sola_2009-2015,LimaSola_2013a,Sola_2013b,Valent2015,Sola_2015_1c,Sola_2015_1a,Sola_2015_1b}).

In the above discussion, the entropic force $F_{rn}$ was calculated from Eq.\ (\ref{eq:Frn}), i.e., $F_{rn} = - T (dS_{rn} / dr_{H})$. Therefore, the heat flow $dQ$ across the horizon can be calculated from $dQ=TdS_{rn}$ as if $S_{rn}$ is exchangeable. Based on this concept and using Eq.\ (\ref{eq:F(rn)}), $dQ$ is given as 
\begin{align}
    dQ  &= T dS = T \left ( \frac{dS}{dr} \right ) dr  =  T  \left ( \frac{dS_{rn}}{dr_{H}}  \right ) dr_{H}          \notag \\
         &=    \gamma  \frac{c^{4}}{G} \left ( \frac{n L_{n} }{2} \right )  r_{H}^{n-2}    dr_{H}            .
\label{eq:dQ=TdS}
\end{align}
Using this heat flow, in Sec.\ \ref{Continuity equation from the first law of thermodynamics}, we derive the continuity equation from the first law of thermodynamics.

\section{Continuity equations}
\label{Two continuity equations}

In this section, two continuity equations are derived from two different methods.
In Sec.\ \ref{Continuity equation from the first law of thermodynamics}, the continuity equation is derived from the first law of thermodynamics. In Sec.\ \ref{Continuity equation from the Friedmann equations}, the continuity equation is derived from the Friedmann and acceleration equations. In the following, irreversible entropy due to matter creation is also considered, i.e., we examine the entropic-force model with matter creation. Accordingly, the formulation discussed here is slightly complicated.

\subsection{Continuity equation from the first law of thermodynamics}
\label{Continuity equation from the first law of thermodynamics}

In this subsection, the continuity equation for the entropic-force model with matter creation is derived from the first law of thermodynamics. For this purpose, the first law of thermodynamics for non-adiabatic processes with matter creation is briefly reviewed, according to the work of Prigogine \textit{et al.} \cite{Prigogine_1988b}.

First, let us consider a closed system containing a constant number of particles $N$ in a volume $V$. From the first law of thermodynamics, the heat flow $dQ$ across a region during a time interval $dt$ is given by 
\begin{equation}
   dQ = dE  + p dV    ,  
\label{eq:ClosedFirstLaw_0}
\end{equation}
where $dE$ and $dV$ are changes in the internal energy $E$ and volume $V$ of the region, respectively \cite{Ryden1}. Dividing this equation by $dt$ gives the following differential form of the first law of thermodynamics \cite{Modak2012}: 
\begin{equation}
   \frac{dQ}{dt} = \frac{dE}{dt}  + p \frac{dV}{dt} = \frac{d}{dt} (\varepsilon    V)  + p \frac{dV}{dt}        ,
\label{eq:ClosedFirstLaw}
\end{equation}
where $\varepsilon$ represents the energy density of cosmological fluids, i.e., $ \varepsilon = \rho c^2$. In addition, $dQ$ is assumed to be related to reversible (exchangeable) entropy $S_{\textrm{rev}}$ \cite{Prigogine_1998}. If adiabatic (and isentropic) processes are considered, i.e., $dQ/dt = 0$, then Eq.\ (\ref{eq:ClosedFirstLaw}) is written as  
\begin{equation}
     \frac{d}{dt} (\varepsilon    V)   + p \frac{dV}{dt}  = 0      .
\label{eq:ClosedFirstLaw(dQ=0)}
\end{equation}
Using Eq.\ (\ref{eq:ClosedFirstLaw(dQ=0)}), the continuity equation for the adiabatic process can be written as  \cite{Ryden1}    
\begin{equation}
       \dot{\rho} + 3  \frac{\dot{a}}{a} \left ( \rho   +  \frac{p}{c^2} \right )  = 0     ,
\label{eq:fluid(dQ=0)}
\end{equation}
where the right-hand side of Eq.\ (\ref{eq:fluid(dQ=0)}) is zero because the right-hand side of Eq.\ (\ref{eq:ClosedFirstLaw(dQ=0)}) is zero.

Now, let us consider more general situations that include matter creation \cite{Lima_1992b,Lima_1992e,Lima_2000,Lima_2014b,Harko_2014,Zimdahl_1993,Lima-Others1996-2008,Lima2010-2014,Lima_Newtonian_1997,Lima2011,Ramos_2014-2014b}. To this end, we assume an open system in which $N$ is time dependent. The matter creation results in the generation of irreversible entropy. For non-adiabatic processes taking place in the open system,  the first law of thermodynamics can be written as 
\begin{equation}
     \frac{d}{dt} (\varepsilon    V)   + p \frac{dV}{dt}  = \left ( \frac{dQ}{dt}  \right )_{\textrm{rev}} + \left ( \frac{\varepsilon + p}{n} \frac{d}{dt} (n V)  \right )_{\textrm{irr}}    ,
\label{eq:OpenFirstLaw_ri}
\end{equation}
where $n$ is the particle number density given by $N/V$ \cite{Prigogine_1988b,Harko_2014}. The entropy per particle is assumed to be constant \cite{Lima_1992b,Lima_2014b}. For details regarding matter creation, see, e.g., Refs. \cite{Prigogine_1988b,Lima_1992b,Harko_2014}. The first term $dQ/dt$ on the right-hand side of Eq.\ (\ref{eq:OpenFirstLaw_ri}) is assumed to be related to reversible entropy $S_{\textrm{rev}}$ due to the exchange (the transfer) of energy. In contrast, the second term on the right-hand side, i.e., $[(\varepsilon + p)/n]  d(nV)/dt$, is related to irreversible entropy $S_{\textrm{irr}}$ due to matter creation. Accordingly, the total entropy change is written as  \cite{Prigogine_1988b}
\begin{equation}
     dS =   dS_{\textrm{rev}} + dS_{\textrm{irr}}     , 
\label{eq:Stotal1}
\end{equation}
with
\begin{equation}
      dS_{\textrm{rev}} = \frac{dQ}{T}      \quad    \textrm{and}    \quad    d S_{\textrm{irr}}  \geq 0    , 
\label{eq:Stotal2}
\end{equation}
where $dS_{\textrm{rev}} = dQ/T$ is assumed \cite{Prigogine_1998}.
Typically, the heat flow $dQ$ is negligible \cite{Prigogine_1988b} when examining adiabatic matter creation \cite{Lima_1992b,Lima_1992e,Lima_2000,Lima_2014b,Harko_2014,Zimdahl_1993,Lima-Others1996-2008,Lima2010-2014,Lima_Newtonian_1997,Lima2011,Ramos_2014-2014b}. 
However, the negligibility should be related to a free parameter $\gamma$ for the temperature, as discussed later.
Accordingly, in this study, we leave the $dQ/dt$ term in Eq.\ (\ref{eq:OpenFirstLaw_ri}).
(Although an entropic-force model in a dissipative universe has been proposed recently, the exchange of energy is neglected \cite{Koma7,Koma8}. More general thermodynamics for matter creation have been discussed by Harko \cite{Harko_2014}.)

To derive the continuity equation, energy flows across the Hubble horizon at $r = r_{H}$ are considered. Therefore, Eq.\ (\ref{eq:OpenFirstLaw_ri}) can be written as  
\begin{align}
   &  \left [ \frac{d}{dt} (\varepsilon    V)   + p \frac{dV}{dt} \right ]_{r=r_{H}}                                                                                                              \notag \\      
   &=  \left [  \left ( \frac{dQ}{dt}  \right )_{\textrm{rev}} + \left ( \frac{\varepsilon + p}{n} \frac{d}{dt} (n V)  \right )_{\textrm{irr}}    \right ]_{r=r_{H}}                   . 
\label{eq:OpenFirstLaw_rev_irr_r=rH}
\end{align}
To calculate the left-hand side of Eq.\ (\ref{eq:OpenFirstLaw_rev_irr_r=rH}), suppose a sphere of arbitrary radius $r$ \cite{Modak2012}. The volume of the sphere is given by $V = 4 \pi r^{3} /3$.  In addition, $r$ is set to be $r_{H}$ after the time derivative in Eq.\ (\ref{eq:OpenFirstLaw_rev_irr_r=rH}) is calculated \cite{Modak2012}.
Concretely speaking, we consider a sphere of arbitrary radius $\hat{r}$ expanding along with the universal expansion:
\begin{equation}
 r(t) = a(t) \hat{r}   .
\end{equation}
The volume $V(t)$ of the sphere is 
\begin{equation}
 V(t) =  \frac{4 \pi}{3} r(t)^3      =  \frac{4 \pi}{3} \hat{r}^3  a(t)^3     .
\label{eq:V(t)}
\end{equation}
From Eq.\ (\ref{eq:V(t)}), the rate of change of the sphere's volume can be given as  \cite{Ryden1,Koma4}
\begin{equation}
\frac{dV}{dt} = \dot{V} = \frac{4 \pi}{3} \hat{r}^3 (3 a^2 \dot{a} ) = V \left ( 3 \frac{\dot{a}}{a} \right )   .
\label{eq:dotV}
\end{equation}
Using Eq.\ (\ref{eq:dotV}), the rate of change of the sphere's internal energy is  
\begin{equation}
\frac{d}{dt} (\varepsilon    V)  =  \dot{\varepsilon} V  + \varepsilon \dot{V} =  \left ( \dot{\varepsilon} + 3 \frac{\dot{a}}{a} \varepsilon \right ) V        .
\label{eq:dotE}
\end{equation}
Substituting Eqs.\ (\ref{eq:dotV}) and (\ref{eq:dotE}) into  $d(\varepsilon  V)/dt + p dV/dt$, and using $\varepsilon = \rho c^{2} $, we have
\begin{align}
\frac{d}{dt} (\varepsilon    V)   + p \frac{dV}{dt} 
            &=      \left ( \dot{\varepsilon} + 3 \frac{\dot{a}}{a} \varepsilon \right ) V  
                                                  + p V \left ( 3 \frac{\dot{a}}{a} \right )                                                       \notag  \\
            &= \left [ \dot{\varepsilon} + 3  \frac{\dot{a}}{a} \left ( \varepsilon   +  p \right )  \right ]  V             \notag  \\           
            &= \left [ \dot{\rho} + 3  \frac{\dot{a}}{a} \left ( \rho   +  \frac{p}{c^2} \right )  \right ] c^2  \left ( \frac{4 \pi}{3} r^3  \right )   .       
\label{eq:dotEpdotV}                            
\end{align}
This equation corresponds to the left-hand side of Eq.\ (\ref{eq:OpenFirstLaw_rev_irr_r=rH}), where the arbitrary radius  $r$ is used.
If we assume adiabatic (and isentropic) processes without dissipation, the right-hand side of Eq.\ (\ref{eq:OpenFirstLaw_rev_irr_r=rH}) is zero. Consequently, the continuity equation is given as $ \dot{\rho} +  3 (\dot{a}/a) ( \rho   +  p/c^2 ) = 0 $.

To calculate the right-hand side of Eq.\ (\ref{eq:OpenFirstLaw_rev_irr_r=rH}), we assume both heat flows related to $S_{\textrm{rev}}$ and matter creation related to $S_{\textrm{irr}}$. In this study, the heat flow can be derived from a general form of entropy [Eq.\ (\ref{eq:S(rn)})]: $ S_{\textrm{rev}} = S_{rn} \propto r_{H}^{n}$ (for $n=2$, $3$, and $4$). Using Eqs.\  (\ref{eq:rH}) and (\ref{eq:dQ=TdS}), the heat flow rate is given as
\begin{align}
\left ( \frac{dQ}{dt}  \right )_{\textrm{rev}} 
                       &=    \gamma  \frac{c^{4}}{G} \left ( \frac{n L_{n} }{2} \right )  r_{H}^{n-2} \frac{dr_{H}}{dt}        \notag  \\
                       &=    \gamma  \frac{c^{4}}{G} \left ( \frac{n L_{n} }{2} \right )  r_{H}^{n-2}  \left ( - \frac{ H  \dot{H} }{ c^2 } r_{H}^3    \right )  \notag  \\ 
                       &=  - \gamma  \frac{c^{2}}{G} \left ( \frac{n L_{n} }{2} \right )  r_{H}^{n+1}  H  \dot{H}    . 
\label{eq:dQdt_rev_FirstTerm}
\end{align}
This equation indicates that the heat flow rate is negligible when $\gamma$ is sufficiently small.
In contrast, the second term on the right-hand side of Eq.\ (\ref{eq:OpenFirstLaw_rev_irr_r=rH}) is related to $S_{\textrm{irr}}$ for matter creation \ \cite{Prigogine_1988b,Prigogine1989,Lima_1992b,Lima_1992e,Lima_2000,Lima_2014b,Harko_2014,Zimdahl_1993,Lima-Others1996-2008,Lima2010-2014,C301,Lima_Newtonian_1997,Lima2011,Ramos_2014-2014b}. Using Eq.\ (\ref{eq:dotE}), and replacing $\varepsilon$ by $n$, we obtain  
\begin{equation}
\frac{d}{dt} (n    V)  =  \left ( \dot{n} + 3 \frac{\dot{a}}{a} n \right ) V        .
\label{eq:dot_n}
\end{equation}
Substituting Eq.\ (\ref{eq:dot_n}) into the second term on the right-hand side of Eq.\ (\ref{eq:OpenFirstLaw_rev_irr_r=rH}), and using $\dot{n} + 3 (\dot{a}/a) n   = n \Gamma$ \cite{Lima_2014b}, we have 
\begin{align}
          \left ( \frac{\varepsilon + p}{n} \frac{d}{dt} (n V)  \right )_{\textrm{irr}} 
                                                                     &=    \frac{\varepsilon + p}{n}  \left ( \dot{n} + 3 \frac{\dot{a}}{a} n \right ) V     \notag  \\ 
                                                                     &=   \frac{\varepsilon + p}{n}  ( n \Gamma  ) V                                                  
                                                                        =    (\varepsilon + p )   \Gamma   V                                                                \notag \\ 
                                                                     &=    \left ( \rho + \frac{p}{c^{2}} \right )  c^{2} V  \Gamma      ,  
\label{eq:SecondTerm}
\end{align}
where $\Gamma$ represents the particle production rate \cite{Harko_2014,Lima_2014b}.

We now calculate Eq.\ (\ref{eq:OpenFirstLaw_rev_irr_r=rH}). Substituting Eqs.\ (\ref{eq:dotEpdotV}), (\ref{eq:dQdt_rev_FirstTerm}), and (\ref{eq:SecondTerm}) into Eq.\ (\ref{eq:OpenFirstLaw_rev_irr_r=rH}), setting $r=r_{H}$, and arranging the resultant equation, we obtain
\begin{align}
       \dot{\rho} + 3  \frac{\dot{a}}{a} \left ( \rho   +  \frac{ p_{e}^{\textrm{irr}} }{c^2} \right )  
        &=  - \gamma  \frac{c^{2}}{G} \left ( \frac{n L_{n} }{2} \right )  r_{H}^{n+1}  H  \dot{H} \frac{1}{\frac{4 \pi}{3} r_{H}^3  c^{2}}   \notag  \\
        &=  - \gamma  \frac{3  n L_{n} }{ 8 \pi G}   r_{H}^{n-2}  H  \dot{H}   \notag  \\
        &=  - \gamma  \frac{3  n L_{n} }{ 8 \pi G}   \left ( \frac{c}{H} \right )^{n-2}  H  \dot{H}   \notag  \\
        &=  - \gamma  \left ( \frac{3  c^{n-2} n L_{n} }{ 8 \pi G}  \right )    H^{3-n}  \dot{H}   , 
\label{eq:fluid_rev_irr}
\end{align}
where $p_{e}^{\textrm{irr}}$ is an effective pressure given by $p_{e}^{\textrm{irr}}  = p + p_{c}^{\textrm{irr}} $, and $p_{c}^{\textrm{irr}}$ is a creation pressure for constant specific entropy in adiabatic matter creation \cite{Prigogine_1988b,Lima_2014b,Harko_2014}. The creation pressure is defined as
\begin{equation}
      p_{c}^{\textrm{irr}} = - \frac{(\rho c^{2} + p) \Gamma}{3H} .
\label{eq:pc_irr}
\end{equation}
For clarity, the effective pressure is written as $p_{e}^{\textrm{irr}}$ because it includes $p_{c}^{\textrm{irr}}$.

In the present paper, a matter-dominated universe (when $p=0$) is considered. Therefore, the effective pressure $p_{e}^{\textrm{irr}}$ is given by $p_{e}^{\textrm{irr}}  = p + p_{c}^{\textrm{irr}} = p_{c}^{\textrm{irr}} $. Consequently, Eqs.\ (\ref{eq:fluid_rev_irr}) and (\ref{eq:pc_irr}) are rewritten as 
\begin{equation}
       \dot{\rho} + 3  \frac{\dot{a}}{a} \left ( \rho   +  \frac{ p_{e}^{\textrm{irr}} }{c^2} \right )  =  \left [ - \gamma  \left ( \frac{3  c^{n-2} n L_{n} }{ 8 \pi G}  \right )    H^{3-n}  \dot{H}    \right ]_{\textrm{rev}}  
\label{eq:fluid_rev_irr_p=0}
\end{equation}
and
\begin{equation}
      p_{e}^{\textrm{irr}} = p_{c}^{\textrm{irr}} = - \frac{ \rho c^{2} \Gamma }{3H} .
\label{eq:pc_irr_p=0}
\end{equation}
Equation\ (\ref{eq:fluid_rev_irr_p=0}) is the modified continuity equation derived from the first law of thermodynamics. The right-hand side of Eq.\ (\ref{eq:fluid_rev_irr_p=0}) is considered to be related to reversible entropy, assuming a general form of entropy on the horizon given by $ S_{\textrm{rev}} = S_{rn} \propto r_{H}^{n}$. In contrast,  $p_{e}^{\textrm{irr}}$  ($= p_{c}^{\textrm{irr}} $) on the left-hand side is related to irreversible entropy due to matter creation \cite{Prigogine_1988b,Lima_2014b,Harko_2014}. 
If the heat flow rate is negligible (i.e., when $\gamma$ is sufficiently small), the continuity equation for adiabatic matter creation is given by  $ \dot{\rho} +  3 (\dot{a}/a) ( \rho   +  p_{e}^{\textrm{irr}}/c^2 ) = 0 $. Substituting $n=2$, $L_{2}=1$, and $p_{e}^{\textrm{irr}} =p$ into Eq.\ (\ref{eq:fluid_rev_irr_p=0}), we obtain the continuity equation discussed in Ref.\ \cite{Koma4}.

\subsection{Continuity equation from the Friedmann and acceleration equations} 
\label{Continuity equation from the Friedmann equations}

In this subsection, the continuity equation is derived from the Friedmann and acceleration equations. We have chosen this route because only two of the three equations are independent \cite{Ryden1}.  To this end, the general Friedmann, acceleration, and continuity equations are reformulated, according to our previous works \cite{Koma4,Koma5,Koma6,Koma7}.
The general equations are applied to the present model, i.e., the entropic-force model with matter creation.

The general Friedmann and acceleration equations for a matter-dominated universe (when $p=0$) are written as
\begin{align}
   H^2     =  \frac{ 8\pi G }{ 3 } \rho       &+ f_{\textrm{rev}}(t)                                                             ,     \label{eq:General_FRW01_f_0}      \\  
  \frac{ \ddot{a} }{ a }  =    -  \frac{ 4\pi G }{ 3 }  \rho   &+ f_{\textrm{rev}}(t)    +  h_{\textrm{irr}}(t)       ,     \label{eq:General_FRW02_g_0}
\end{align}
with 
\begin{equation}
 f_{\textrm{rev}}(t)   \geq 0  \quad \textrm{and}  \quad   h_{\textrm{irr}}(t)   \geq 0   , 
\label{eq:g(t)_GE_f(t)_00}
\end{equation}
where $f_{\textrm{rev}}(t)$ and $h_{\textrm{irr}}(t)$ are general extra driving-terms \cite{Koma7}. In this formulation, $f_{\textrm{rev}}(t)$ is considered to be related to reversible entropy $S_{\textrm{rev}}$, whereas  $h_{\textrm{irr}}(t)$ is related to irreversible entropy $S_{\textrm{irr}}$. Consequently, Eq.\ (\ref{eq:General_FRW02_g_0}) can be rearranged as
\begin{equation}
  \frac{ \ddot{a} }{ a }    =  -  \frac{ 4\pi G }{ 3 } \left ( \rho + \frac{3 p_{e}^{\textrm{irr}} }{c^2}  \right ) + f_{\textrm{rev}}(t)                                       ,
\label{eq:General_FRW02_g_f}
\end{equation}
where the effective pressure $p_{e}^{\textrm{irr}}$ is given by
\begin{equation}
  p_{e}^{\textrm{irr}} \equiv  - \frac{  c^{2}  h_{\textrm{irr}}(t)  } {  4\pi G }  .
\label{eq:General_pe}
\end{equation}
In a matter-dominated universe (when $p=0$), the effective pressure $p_{e}^{\textrm{irr}}$ is given by $p_{e}^{\textrm{irr}}  = p + p_{c}^{\textrm{irr}} = p_{c}^{\textrm{irr}}$. Here, $p_{c}^{\textrm{irr}}$ is interpreted as a pressure derived from $S_{\textrm{irr}}$. Accordingly, $p_{e}^{\textrm{irr}}$ is equivalent to that in the previous subsection because the same matter creation is assumed.

We now calculate the general continuity equation \cite{Koma4,Koma5,Koma6} from the general Friedmann and acceleration equations. The general continuity equation in a matter-dominated universe becomes
\begin{equation}
       \dot{\rho} + 3  \frac{\dot{a}}{a}  \rho   =  \frac{3}{4 \pi G} H \left(  h_{\textrm{irr}}(t)  -  \frac{\dot{f}_{\textrm{rev}}(t)}{2 H }      \right )                   .
\label{eq:drho_General_00}
\end{equation}
This equation can be rewritten as 
\begin{equation}
       \dot{\rho} + 3  \frac{\dot{a}}{a}  \rho   =  \rho  \Gamma_{\textrm{irr}}   - \Theta_{\textrm{rev}}  ,
\label{eq:drho_General_01}
\end{equation}
or equivalently
\begin{equation}
       \dot{\rho} + 3  \frac{\dot{a}}{a} \left ( \rho + \frac{p_{e}^{\textrm{irr}}}{c^{2}}  \right )  =   -\Theta_{\textrm{rev}}   ,
\label{eq:drho_General_011}
\end{equation}
where, using $p_{e}^{\textrm{irr}}$ from Eq.\ (\ref{eq:General_pe}),  $\Gamma_{\textrm{irr}}$ is given by
\begin{equation} 
    \Gamma_{\textrm{irr}}  = \frac{3 H}{4 \pi G}  \frac{ h_{\textrm{irr}}(t)}{ \rho } = - 3 H \frac{ p_{e}^{\textrm{irr}} }{\rho c^{2}} , 
\label{eq:Gamma_0}
\end{equation}
and $\Theta_{\textrm{rev}}$ is defined by 
\begin{equation} 
        \Theta_{\textrm{rev}}  =   \frac{3}{8 \pi G}   \dot{f}_{\textrm{rev}}(t)   .
\label{eq:Q_0}
\end{equation}
$\Gamma_{\textrm{irr}}$ in Eq.\ (\ref{eq:Gamma_0}) is equivalent to $\Gamma$ in Eq.\ (\ref{eq:pc_irr_p=0}).
That is, the general function $ h_{\textrm{irr}}(t)$ is a constant given by
\begin{equation}
  h_{\textrm{irr}}(t) = h_{\textrm{irr}0} =  -   \frac{ 4\pi G  p_{e}^{\textrm{irr}} }{  c^{2}  }  = \textrm{constant}     .
\label{eq:h_irr}
\end{equation}
On the other hand, from Eqs.\ (\ref{eq:FRW2_(rn)}) and (\ref{eq:FRW1_(rn)}), $f_{\textrm{rev}}(t)$ can be written as 
\begin{equation}
  f_{\textrm{rev}}(t) =     \gamma  \left ( \frac{c^{n-2}  n L_{n} }{  2 }  \right  )  H^{4-n}      \quad (n=2,3,4), 
\label{eq:f_rev}
\end{equation}
where $n=2$, $3$, and $4$ correspond to indices  of area, volume, and quartic entropies, respectively. Accordingly, the Friedmann and acceleration equations for the present model are summarized as
\begin{equation}
   H^2     =  \frac{ 8\pi G }{ 3 } \rho  + \gamma  \left ( \frac{c^{n-2}  n L_{n} }{  2 }  \right  )  H^{4-n}      ,
\label{eq:FRW01_EntroMatter}
\end{equation}
\begin{equation}
  \frac{ \ddot{a} }{ a }    =  -  \frac{ 4\pi G }{ 3 } \left ( \rho + \frac{3 p_{e}^{\textrm{irr}} }{c^2}  \right ) +  \gamma  \left ( \frac{c^{n-2}  n L_{n} }{  2 }  \right  )  H^{4-n}      , 
\label{eq:FRW02_EntroMatter}
\end{equation}
where the last term on the right-hand side is the entropic-force term derived from a general form of entropy on the horizon.

Now, we calculate $\Theta_{\textrm{rev}}$ on the right-hand side of Eq.\ (\ref{eq:drho_General_011}). Substituting Eq.\ (\ref{eq:f_rev}) into Eq.\ (\ref{eq:Q_0}), and rearranging the resultant equation, we obtain
\begin{align} 
       \Theta_{\textrm{rev}}   &=   \frac{3}{8 \pi G} \times \gamma  \left ( \frac{c^{n-2}  n L_{n} }{  2 }  \right  ) (4-n) H^{3-n} \dot{H}    \notag \\
                                         &=   \gamma \left ( \frac{3 c^{n-2}  n L_{n}}{8 \pi G}  \right  )    \left ( \frac{4-n}{  2 }  \right  )  H^{3-n} \dot{H}         .
\label{eq:Q_10}
\end{align}
In addition, substituting Eq.\ (\ref{eq:Q_10}) into Eq.\ (\ref{eq:drho_General_011}), we have 
\begin{align} 
    &   \dot{\rho} + 3  \frac{\dot{a}}{a} \left ( \rho + \frac{p_{e}^{\textrm{irr}}}{c^{2}}  \right )                                             \notag \\
    &= \left [  - \gamma \left ( \frac{3 c^{n-2}  n L_{n}}{8 \pi G}  \right  )    \left ( \frac{4-n}{  2 }  \right  )  H^{3-n} \dot{H}     \right ]_{\textrm{rev}}   .
\label{eq:drho_General_011_entropic}
\end{align}
This equation is the modified continuity equation for the entropic-force model with matter creation, which is derived from the Friedmann and acceleration equations.
The right-hand side of Eq.\ (\ref{eq:drho_General_011_entropic}) depends on the general form of entropy on the horizon. In this way, the two continuity equations for the present model are derived from the different methods. In the next section, the consistency of the two continuity equations is discussed. (The present model is considered to be a kind of $\Lambda (t)$CDM model in a dissipative universe. Brevik \textit{et al.} have recently examined a similar cosmological system with two interacting fluids in a dissipative universe \cite{Brevik_2015a}.)

Note that, as shown in Eq.\ (\ref{eq:f_rev}), in order to derive the continuity equation, we assume that $f_{\textrm{rev}}(t)$ is an entropic-force term.  Additionally, in the previous subsection, the continuity equation was derived from the first law of thermodynamics, assuming $ S_{\textrm{rev}} = S_{rn} \propto r_{H}^{n}$.  Accordingly, it may seem that the exchangeability of the entropy on the horizon is assumed beforehand. Therefore, the validity should be confirmed by studying the consistency of the two continuity equations, as discussed in the next section.

\section{Consistency of the two continuity equations} 
\label{Consistency}

In Sec.\ \ref{Two continuity equations}, two continuity equations for the entropic-force model with matter creation were derived using different methods. In this section, we examine the consistency of those two continuity equations. Moreover, we discuss the exchangeability of the entropy on the horizon of the universe in entropic cosmology. If the two continuity equations agree, we can interpret the agreement as a sign that the entropy behaves exchangeably. Note that we admit the possibility that the consistency is not directly related to the exchangeability.

To study the consistency of two continuity equations, the two equations are written again. From Eq.\ (\ref{eq:fluid_rev_irr_p=0}), the continuity equation derived from the first law of thermodynamics is written as 
\begin{equation}
       \dot{\rho} + 3  \frac{\dot{a}}{a} \left ( \rho   +  \frac{ p_{e}^{\textrm{irr}} }{c^2} \right )  =  - \gamma  \left ( \frac{3  c^{n-2} n L_{n} }{ 8 \pi G}  \right )    H^{3-n}  \dot{H}   , 
\label{eq:fluid_rev_irr_p=0_2}
\end{equation}
where $n$ is an index of entropy: i.e., $n=2$, $3$, and $4$ correspond to indices of the area, volume, and quartic entropies, respectively. In contrast, from Eq.\ (\ref{eq:drho_General_011_entropic}), the continuity equation derived from the Friedmann and acceleration equations is 
\begin{equation}
       \dot{\rho} + 3  \frac{\dot{a}}{a} \left ( \rho + \frac{p_{e}^{\textrm{irr}}}{c^{2}}  \right )  =  - \gamma \left ( \frac{3 c^{n-2}  n L_{n}}{8 \pi G}  \right  )    \left ( \frac{4-n}{  2 }  \right  )  H^{3-n} \dot{H}    .
\label{eq:drho_General_0110_entropic}
\end{equation}
As shown in Eqs.\ (\ref{eq:fluid_rev_irr_p=0_2}) and (\ref{eq:drho_General_0110_entropic}), the two left-hand sides agree because an equivalent matter-creation is assumed. Interestingly, the two right-hand sides are also consistent with each other, except for the coefficient $(4-n)/2$ in Eq.\ (\ref{eq:drho_General_0110_entropic}). Therefore, the two right-hand sides are in absolute agreement when $n=2$, which corresponds to an area entropy. A similar non-zero right-hand side appears in energy exchange cosmology \cite{Koma4}. This consistency of the two continuity equations may imply the exchange (the transfer) of energy in entropic cosmology. For example, the interchange of energy between the bulk (the universe) and the boundary (the horizon of the universe) \cite{Lepe1} is a viable scenario from the viewpoint of the holographic principle. Of course, strictly speaking, the two right-hand sides are slightly different when $n=3$ and $4$ due to the coefficient $(4-n)/2$ in Eq.\ (\ref{eq:drho_General_0110_entropic}). In this case, the entropic-force model should be considered to be a type of energy exchange cosmology between dark matter and effective dark energy \cite{Sola_2014a}. When $n > 4$ (corresponding to higher-order forms of entropy), the two right-hand sides have opposite sign due to the coefficient $(4-n)/2$. This opposite sign may be interpreted as a sign that the direction of heat flows could be reversed when Eq.\ (\ref{eq:fluid_rev_irr_p=0_2}) is derived. Alternatively, these results simply imply that the Bekenstein entropy (an area entropy) is the most suitable for describing entropic cosmology.

The coefficient, $(4-n)/2$, plays an important role because it affects the difference between the two continuity equations. The interpretation of the coefficient should provide new insights into the exchange of energy in entropic cosmology although it has not yet been made clear.  Note that we can obtain an effective continuity equation similar to bulk viscous and CCDM models by moving the non-zero right-hand side to the other side and extending an effective description for the pressure.

In this short paper, we have focused on the consistency of the two continuity equations. Therefore, we do not examine the properties of the present model in detail.  However, it is possible to evaluate them roughly because the cosmological equations used here are similar to those of the $\Lambda(t)$CDM, CCDM, and entropic-force models \cite{Valent2015,Lima2011,Sola_2014a,Koma6}. For example, background evolutions of the universe are essentially equivalent to those described by an extended entropic-force model in Ref.\ \cite{Koma6}. In addition, a unified formulation for density perturbations \cite{Koma6} can be applied to the present model when $f_{\textrm{rev}}(t)=0$ or $h_{\textrm{irr}}(t)=0$. Accordingly, in the next paragraph, we briefly discuss the properties of the present model, according to the previous studies \cite{Basilakos1,Sola_2014a,Valent2015,Koma4,Koma5,Koma6,Lima2011,Ramos_2014-2014b}.

As shown in Eqs.\ (\ref{eq:FRW01_EntroMatter}) and (\ref{eq:FRW02_EntroMatter}), the Friedmann and acceleration equations include $H^{4-n}$ terms related to $f(t)_{\textrm{rev}}$. Moreover, the acceleration equation includes the effective pressure $p_{e}^{\textrm{irr}}$ related to the constant $h_{\textrm{irr}0}$ term for matter creation. First, we focus on the entropic-force term $H^{4-n}$. In fact, the entropic-force model, which includes each of the $H^{2}$, $H$, and constant terms, agrees well with observed supernova data \cite{Koma4,Koma5,Koma6}. That is, each of the three terms can properly describe the accelerated expansion of the late universe. However, Basilakos and Sol\`{a} have shown that simple combinations of pure Hubble terms, i.e., $H^{2}$, $\dot{H}$, and $H$ terms, are insufficient for a complete description of the growth rate for clustering related to structure formations \cite{Sola_2014a}. Similarly, several combinations of $H^{2}$, $\dot{H}$, $H$, and constant terms in the $\Lambda (t)$CDM model have been examined by G\'{o}mez-Valent \textit{et al.} \cite{Valent2015}. Those studies indicate that the constant term plays an important role in the discussion of observations of cosmological fluctuations \cite{Basilakos1,Sola_2014a,Valent2015,Koma6}. In the $\Lambda(t)$CDM model, such a constant term is obtained from an integral constant of the renormalization group equation for the vacuum energy density \cite{Sola_2013b}. A similar constant term (corresponding to $h_{\textrm{irr}0}$) appears in CCDM models. However, in CCDM models, a negative sound speed \cite{Lima2011} and the existence of clustered matter \cite{Ramos_2014-2014b} are necessary to properly describe the growth rate \cite{Koma8}. Therefore, the entropic-force model with a non-zero $h_{\textrm{irr}}(t)$ term (not only the constant term but also the $H$ term) is inconsistent with the observed growth rate, especially for a low redshift, as we have previously shown  \cite{Koma6}. We have also shown that a weakly dissipative model (similar to the $\Lambda$CDM model) describes observations of the cosmic microwave background radiation temperature, whereas a strong dissipative model (similar to the CCDM model) does not \cite{Koma8}.

The present study and previous studies imply that an area entropy (which leads to $H^{2}$ terms), a constant term, and a weakly dissipative universe are favored.
Accordingly, $ f_{\textrm{rev}}(t) = C_{0} + C_{1} H^{2}$ and $h_{\textrm{irr}}(t) = 0$ can be proposed for one of the favored models, where $C_{0}$ and $C_{1}$ are constants. The favored model can be interpreted as a particular case of $\Lambda (t)$CDM models. This type of $\Lambda(t)$CDM model has been recently examined by, e.g., Lima \textit{et al.} \cite{LimaSola_2013a}, G\'{o}mez-Valent \textit{et al.} \cite{Valent2015}, and Basilakos and Sol\`{a} \cite{Basilakos1}.

From Eqs.\ (\ref{eq:fluid_rev_irr_p=0_2}) and (\ref{eq:drho_General_0110_entropic}), the two continuity equations are found to be slightly inconsistent with each other when $n \neq 2$. 
Interestingly, when $n \neq 2$, the maximum tension principle does not work for generalized entropic-force models as well, as recently examined by D\c{a}browski and Gohar \cite{Gohar_2015a}.
That is, $n=2$ is suitable both for the consistency of the two continuity equations and for the maximum tension principle.
The results imply that the entropic-force model alone may be difficult to solve the cosmological constant problem because an additive constant term is obtained not from $n=2$ but from $n = 4$. 
In addition, as mentioned previously, it is difficult to properly describe not only a decelerating and accelerating universe but also structure formations, without adding the constant term \cite{Basilakos1,Sola_2014a,Sola_2009-2015,LimaSola_2013a,Sola_2013b,Valent2015,Koma6,Gohar_2015b}. 
To solve these difficulties, the entropic-force model for $n=2$ should be appropriately coupled with $\Lambda (t)$CDM models.
In particular, the favored model proposed in the above paragraph is expected to play an important role theoretically and phenomenologically.

In fact, recent studies imply that $\Lambda (t)$CDM models based on power series of the Hubble rate are likely more suitable both for a theoretical explanation and for a phenomenological description than the standard $\Lambda$CDM model.
See, e.g., Ref.\ \cite{Sola_2015_1c} and the third paper of Ref.\ \cite{Valent2015}.
Matter is conserved in the $\Lambda (t)$CDM models.
The present entropic-force model is expected to be related to the $\Lambda (t)$CDM models. 
However, it should be difficult to distinguish the entropic-force model from the $\Lambda (t)$CDM model in practice, when the formulations are the same and dissipative terms are zero, i.e., $h_{\textrm{irr}}(t) = 0$. 
Of course, when $h_{\textrm{irr}}(t) \neq 0$, we can distinguish between the two models even if background evolutions of the universe are the same. 
However, our previous studies imply that a weakly dissipative universe is favored \cite{Koma6,Koma7,Koma8}.
(A slowly varying gravitational coupling is assumed for the $\Lambda (t)$CDM model examined in Ref.\ \cite{Sola_2015_1c}, unlike for the present entropic-force model.
Accordingly, it may be possible to distinguish the entropic-force model for $h_{\textrm{irr}}(t) = 0$ from the $\Lambda (t)$CDM model, if a varying gravitational constant is revealed through observations.)

Finally, the inflation of the early universe in entropic cosmology is briefly discussed. 
In the present study, $H^{4-n}$ terms are obtained from entropic forces. 
Accordingly, the exponent $4-n$ decreases with $n$. 
However, higher exponents should be required for the inflation. 
The higher exponent cannot be obtained from the present entropic force, without assuming $n \leq 0$.
Probably, this problem can be solved by introducing logarithmic entropic corrections which generate $H^4$ terms (see, e.g., the second paper of Ref.\ \cite{Easson12}).
Such an entropic-force model can be interpreted as a particular case of $\Lambda (t)$CDM models as well. 
For example, not only $H^{4}$ terms \cite{Sola_2015_1a} but also $H^{m}$ terms \cite{Sola_2015_1b} (corresponding to an arbitrary exponent of $H$) have been recently examined in $\Lambda (t)$CDM models.
To acquire a deeper understanding of cosmology, we need to study general relativity from various viewpoints \cite{Antoci,Bousso_2015,Sola_2016}.

\section{Conclusions}
\label{Conclusions}

Entropic-force models assume several forms of entropy on the horizon of the universe, where the entropy can be considered to behave as if it were exchangeable. To study the consequences of this assumption, a phenomenological entropic-force model has been formulated, focusing on a homogeneous, isotropic, spatially flat, and matter-dominated universe. For this formulation, a general form of entropy proportional to the $n$-th power of the Hubble horizon, i.e., $S_{rn} \propto r_{H}^{n}$, is used. Here, the Bekenstein entropy (an area entropy), the Tsallis--Cirto black-hole entropy (a volume entropy), and a quartic entropy are represented by $n=2$, $3$, and $4$, respectively. Consequently, $H^{4-n}$ terms for the Friedmann and acceleration equations are obtained from entropic-forces. That is, the $H^{2}$, $H$, and constant entropic-force terms are confirmed to be systematically derived from the area, volume, and quartic entropies, respectively.

In addition, irreversible entropy due to matter creation has been included in the current formulation. Using the present model, we have examined whether the entropy $S_{rn}$ on the horizon behaves exchangeably or not. To this end, two continuity equations for the present model are derived from two different methods. The first continuity equation is derived from the first law of thermodynamics, whereas the second equation is derived from the Friedmann and acceleration equations. The two continuity equations are found to be consistent with each other. In particular, the two equations agree completely when $n=2$, which corresponds to the Bekenstein entropy. This consistency may imply the exchange (the transfer) of energy in entropic cosmology. The interchange of energy between the bulk (the universe) and the boundary (the horizon of the universe) is a viable scenario consistent with the holographic principle. Alternatively, the entropy on the horizon in the entropic-force model can be interpreted as an effective dark energy. The present study should provide new insights into the entropic, energy-exchange, and time-varying $\Lambda(t)$ cosmologies and bridge the gap between them.

\begin{acknowledgements}
The authors wish to thank the anonymous referee and H. Gohar for very valuable comments which improve the paper. 
\end{acknowledgements}

\end{document}